\documentclass[a4paper,oneside]{article}

\usepackage{microtype}
\usepackage[sc]{mathpazo}
\usepackage[T1]{fontenc}
\usepackage[DIV=11,BCOR=0cm,headinclude=true]{typearea}

\usepackage{amsfonts}
\usepackage{amsmath}
\usepackage{amssymb}
\usepackage{amsthm}
\usepackage{graphicx}
\usepackage{float}
\usepackage[config,labelfont=sc,textfont=sl]{caption,subfig}
\usepackage[british]{babel}

\usepackage[maxbibnames=100,autocite=superscript,style=custom-nature,sorting=none,defernumbers=true]{biblatex}

\usepackage{authblk}

\newcommand{\bv}[1]{\boldsymbol{#1}}

\DeclareMathOperator*{\diag}{diag}
\DeclareMathOperator*{\real}{Re}
\DeclareMathOperator*{\imag}{Im}

\author[1]{Parry Y.\ Chen}
\author[1]{Jacob Ben-Yakar}
\author[2]{Yonatan Sivan}
\affil[1]{School of Physics and Astronomy, Raymond and Beverly Sackler Faculty of Exact Sciences, Tel Aviv University, Israel}
\affil[2]{Unit of Electro-optic Engineering, Ben-Gurion University, Israel}

\title{Reinterpreting the magnetoelectric coupling of polarizability tensors of infinite cylinders using symmetry: a simple TM/TE view}
\date{24th May 2016}

\begin{document}
\maketitle

\begin{abstract}
Recently, Strickland et al.\ retrieved dynamic polarizabilities of infinitely long wires at oblique incidence, reporting non-zero magnetoelectric coupling, seemingly defying existing theorems which forbid this in centrosymmetric scatterers. We reconcile this finding with existing symmetry restrictions on microscopic polarizabilities using a property of line dipoles. This motivates a reformulation of cylinder polarizability, yielding diagonal tensors that decompose the response into TM and TE contributions, simplifying subsequent treatment by homogenization theories. A transformation is derived between Strickland et al.'s formulation and our reformulation, allowing magnetoelectric coupling to be identified as the contrast between TM and TE responses, and enabling simple geometric insights into all its scaling and symmetry properties.
\end{abstract}

\section{Introduction}
Since the advent of metamaterials, the long thin cylinder has been a fundamental building block in a multitude of metamaterial designs.\autocite{pendry1996extremely} It features prominently in bulk metamaterial designs, often consisting entirely of long circular cylinders, either metallic or dielectric, arranged on a periodic lattice.\autocite{simovski2012wire} These include all-dielectric negative index metamaterials based on Mie resonances\autocite{obrien2002photonic,peng2007experimental,schuller2007dielectric,zhao2009mie,vynck2009all,paniagua-dominguez2013ultra} and dark modes,\autocite{jain2014large} hyperbolic media,\autocite{smith2003electromagnetic,lu2008superlens,simovski2013optimization,poddubny2013hyperbolic} and drawn metamaterial fibers.\autocite{tuniz2010weaving,yaman2011arrays,tao2015multimateriala} The scope of applications is similarly broad, encompassing super-resolution endoscopes,\autocite{belov2005canalization,silveirinha2006nonlocal,shvets2007guiding} planar superlenses,\autocite{lu2008superlens,liu2008all,lemoult2012polychromatic} enhanced coupling to quantum sources,\autocite{yao2011three,jacob2012broadband,cortes2012quantum} solar collectors,\autocite{simovski2013optimization,law2005nanowire,fan2009three} and single-molecule bio-sensors.\autocite{kabashin2009plasmonic} Applications also span the electromagnetic spectrum, relating even to propagation of radio waves through forests.\autocite{li2010investigation}

Arrays of long thin cylinders also appear on metasurfaces, both as structures fabricated parallel to a planar substrate or etched into the substrate itself. Through geometry alone, extensive engineering of reflection phases and angles is possible.\autocite{zhang2013infrared,pors2013plasmonic,du2013nearly,yanik2009hybridized} Beyond periodic arrays, cylinders have been arranged in linear chains and clusters, enabling waveguiding and antenna functionalities.\autocite{smith2003electromagnetic} Single cylinder designs, especially core-cladding designs, use the interplay between electric and magnetic Mie resonances to shape the profile of scattered and emitted light,\autocite{novotny2011antennas,liu2013scattering} with application to cloaking,\autocite{silveirinha2008cloaking} microscopy,\autocite{schroter2001surface} and photodectection.\autocite{fan2012invisible} The long thin cylinder has a vast range of applications, and the cross-section presented here is by no means exhaustive.

Retrieving the response of individual cylinders is a fundamental step towards characterizing the effective electromagnetic response of a metamaterial. For optically thin cylinders, scattering is adequately described by only dipolar fields, considerably simplifying subsequent treatment using effective medium theories.\autocite{raab2004multipole,silveirinha2006nonlocal,alu2011first} Analytical results are available, enabling a systematic design process. This proceeds analogously to the atomic polarizability, which quantifies the distortion of electron clouds due to impinging fields, ultimately yielding the macroscopic constitutive relations. Polarizability $\bar{\bar{\alpha}}$ is defined as
\begin{align}
\begin{bmatrix} \bv{p} \\ \bv{m}  \end{bmatrix} &=
\bar{\bar{\alpha}}
\begin{bmatrix}
\bv{E^\textrm{inc}} \\
\bv{H^\textrm{inc}}
\end{bmatrix},
&
\bar{\bar{\alpha}} &=
\begin{bmatrix}
\bar{\bar{\alpha}}^{ee} & \bar{\bar{\alpha}}^{em}\\ 
\bar{\bar{\alpha}}^{me} & \bar{\bar{\alpha}}^{mm}
\end{bmatrix},
\label{eq:alpha}
\end{align} 
where $\bv{p}$ and $\bv{m}$ are the electric and magnetic dipole moments induced by incident fields. The diagonal blocks of $\bar{\bar{\alpha}}$ describe the electric and magnetic polarizabilities of a scatterer. More generally, a scatterer may have non-zero off-diagonal blocks, $\bar{\bar{\alpha}}^{em}$ and $\bar{\bar{\alpha}}^{me}$, corresponding to magnetoelectric coupling whereby magnetic and electric fields induce electric and magnetic dipole moments, respectively. In macroscopic constitutive relations, these are commonly defined as $\bv{D} = \bar{\bar{\epsilon}}\bv{E} + \bar{\bar{\xi}}\bv{H}$, and $\bv{B} = \bar{\bar{\zeta}}\bv{E} + \bar{\bar{\mu}}\bv{H}$. For the infinite cylinder, the polarizability tensor depends on the angle of incidence due to the longitudinal translational symmetry, which leads to constitutive relations that also depend on the longitudinal propagation constant.\autocite{silveirinha2006nonlocal,belov2003strong}

Due to its fundamental importance, a number of recent papers retrieve $\bar{\bar{\alpha}}$ for infinitely long cylinders under specific incidence conditions, such as normal incidence.\autocite{schroter2001surface,silveirinha2006nonlocal,vynck2009all,kallos2012resonance} And even though scattering from infinite cylinders is a thoroughly investigated textbook problem,\autocite{bohren1983absorption} explicit expressions for the full polarizability tensor $\bar{\bar{\alpha}}$ at oblique incidence were only published recently by Strickland et al.\autocite{strickland2015dynamic} Surprisingly though, a magnetoelectric coupling was reported. Since $\bv{E}$ is odd while $\bv{H}$ is even under inversion, magnetoelectric coupling is considered forbidden in structures containing a center of inversion. Furthermore, local constitutive relations of 2D lattices of such infinite wires after homogenization are known to have zero magnetoelectric coupling.\autocite{silveirinha2007metamaterial,reyes-avendano2011photonic} Strickland et al.\ ascribe the unexpected magnetoelectric coupling to the asymmetric nature of oblique incidence, likening it to the observation of optical activity in achiral split ring resonators, termed pseudochirality or extrinsic chirality.\autocite{saadoun1992reciprocal,tretyakov1993eigenwaves,plum2009extrinsic}

Symmetry restrictions on magnetoelectric coupling have been systematically studied in the field of solid state physics,\autocite{birss1964symmetry,odell1970electrodynamics} and more recently in metamaterials for split ring resonators.\autocite{marques2002role,padilla2007group,baena2007systematic} The result is general: no magnetoelectric coupling may exist in the microscopic polarizability $\bar{\bar{\alpha}}$ of a scatterer with both temporal and inversion symmetry.\autocite{barron2004molecular,raab2004multipole} Analogous symmetry prohibitions exist for macroscopic constitutive relations, requiring $\bar{\bar{\zeta}}(\omega) = \bar{\bar{\xi}}(\omega) = 0$ for inversion symmetric structures.\autocite{silveirinha2007metamaterial,reyes-avendano2011photonic} Equivalently, macroscopic magnetoelectric coupling can be described by first order spatial dispersion,\autocite{landau1984electrodynamics} by incorporating magnetic responses into a permittivity tensor both temporally and spatially dispersive, $\bv{D} = \bar{\bar{\epsilon}}(\omega,\bv{k}) \bv{E}$. Weak spatial dispersion permits expansion in powers of $\bv{k}$, 
\begin{equation}
\epsilon_{ij}(\omega, \bv{k}) = \epsilon_{ij}(\omega) + i\gamma_{ijk}(\omega)k_k + \beta_{ijkl}(\omega)k_kk_l + \cdots
\end{equation}
where $\gamma_{ijk}(\omega)$ corresponds precisely to non-spatially dispersive $\bar{\bar{\zeta}}(\omega)$ and $\bar{\bar{\xi}}(\omega)$. Identical symmetry constraints apply, requiring $\gamma_{ijk}(\omega)$ to vanish.\autocite{landau1984electrodynamics} 

More recently, homogenized magnetoelectric coupling tensors which are themselves $\bv{k}$-dependent, $\bar{\bar{\zeta}}(\omega, \bv{k})$ and $\bar{\bar{\xi}}(\omega, \bv{k})$, have been studied. This arises when a medium is excited by an arbitrary excitation $(\omega, \bv{k})$, as opposed to homogenization based on excitation-free eigenmodes of a medium along its dispersion relation $\omega(\bv{k})$.\autocite{alu2011first,li2007non,fietz2010current,alu2011restoring} Only in this context was it shown that even centrosymmetric metamaterials, such as spheres on a cubic lattice, can have non-zero $\bar{\bar{\zeta}}(\omega, \bv{k})$ and $\bar{\bar{\xi}}(\omega, \bv{k})$. This magnetoelectric coupling emerges during the homogenization procedure, due to the phase delay of the excitation between adjacent unit cells.\autocite{alu2011first} Since such arguments pertain to macroscopic constitutive relations derived from the homogenization of a lattice, their applicability to the microscopic polarizability $\bar{\bar{\alpha}}$ of an individual cylinder is not immediately apparent.

In this paper, we delve into the origins of the non-zero magnetoelectric coupling of infinite cylinders at the level of microscopic polarizability $\bar{\bar{\alpha}}$. We present two independent arguments, serving two different purposes. Firstly, we affirm the results of Strickland et al.\ by providing formal symmetry arguments to reconcile the unexpected magnetoelectric coupling with existing symmetry theorems. Secondly, we derive a transformation which reformulates the magnetoelectric tensor $\bar{\bar{\alpha}}$, restoring the expected diagonal response by partitioning the response into its TM and TE components. This complementary and equivalent formulation predicts identical results, but its diagonal form enables simpler homogenization procedures, and provides simple geometric interpretations for the magnetoelectric coupling and its behavior.

In Section \ref{sec:group} we discuss symmetry, demonstrating how magnetoelectric coupling arises despite inversion symmetry, stemming from the intrinsic longitudinal $k$-variation of line dipoles which distinguishes them from point dipoles. We use group representation theory, a systematic and universal formalism for treating symmetry to derive all the symmetry restrictions on $\bar{\bar{\alpha}}$, including its symmetry-forbidden elements. Using symmetry arguments alone, we exactly reproduce the structure $\bar{\bar{\alpha}}$ derived by Strickland et al. This provides justification for the magnetoelectric coupling of $\bar{\bar{\alpha}}$ at the level of individual cylinders, without the need for analogies with macroscopic constitutive relations derived from the homogenization of a lattice. 

However, our symmetry analysis also reveals that magnetoelectric coupling is not a necessary consequence of the $k$-dependence of line dipoles. In Section \ref{sec:trans}, we develop an alternative formulation of $\bar{\bar{\alpha}}$ that obviates the need for magnetoelectric coupling terms. Our key insight is that the simple decomposition of $\bar{\bar{\alpha}}$ into its TM and TE contributions yields diagonal polarizabilities with zero magnetoelectric coupling. The equivalence of the TM/TE reformulation is established by a mathematical transformation which we derive. Simpler, alternative insight into the physical origins of the magnetoelectric coupling is enabled, bypassing the need to invoke the formal group-theoretic symmetry arguments of Section \ref{sec:group}, or rely on macroscopic analogies used by Strickland et al.\ stemming from the relatively novel and unfamiliar form of spatial dispersion $\bar{\bar{\zeta}}(\omega,\bv{k})$ and $\bar{\bar{\xi}}(\omega,\bv{k})$. Such abstract arguments compound the counter-intuitive nature of the magnetoelectric coupling, rendering it prone to misinterpretation. 

In Section \ref{sec:discuss}, we employ our reformulation to demonstrate that the off-diagonal magnetoelectric coupling terms account for the difference between TM and TE responses, allowing two dissimilar diagonal responses to be combined onto the single tensor $\bar{\bar{\alpha}}$. In the process, we dispel some potential misconceptions that surround the existence of the magnetoelectric coupling. In particular, magnetoelectric coupling is not a consequence of the cross-coupling between TM and TE polarized plane waves during scattering at oblique incidence, but requires only that the TM and TE responses differ. 

All magnetoelectric properties have simple reinterpretations based on the geometry of plane waves when decomposed into the TM/TE view. Quantitative properties are discussed in Section \ref{sec:numerics}. We show that the magnetoelectric coupling of dielectric cylinders exhibits weak quartic scaling at long wavelengths, but becomes prominent at Mie resonances. We also show that perfect electric conductors exhibit a stronger quadratic scaling at long wavelengths, despite having zero polarization cross-coupling even for oblique incidence scattering. Then, symmetry properties are discussed in Section \ref{sec:symmetry}, such as the odd dependence on wavevector. Its geometric origin becomes apparent once the role of the magnetoelectric coupling as a proxy is identified. We reveal one practical consequence of this odd symmetry, showing that the magnetoelectric coupling of infinite cylinders, not necessarily of circular cross-sections, cannot be directly retrieved using numerical schemes that employ counterpropagating waves, but instead can be indirectly obtained using our transformation.

\section{Symmetry Restrictions}
\label{sec:group}
Since symmetry properties of polarizability tensors are predominantly phrased in the language of group theory, we establish the consistency of the magnetoelectric formulation of Strickland et al.\ with the extensive existing literature, and show the simple extension of group theory techniques from point dipoles to line dipoles.

Consider an infinite circular cylinder modeled by line dipoles under impinging radiation with harmonic spatial variation $e^{i\beta z}$ along its axis. Define electric and magnetic line dipole moments concentric with the cylinder axis with said $e^{i\beta z}$ variation, 
\begin{equation}
\begin{aligned}
\bv{p}_z &= p_z \bv{\hat{z}} e^{i\beta z},& \quad
\bv{p}_x &= p_x \bv{\hat{x}} e^{i\beta z},& \quad
\bv{p}_y &= p_y \bv{\hat{y}} e^{i\beta z},
\\ 
\bv{m}_z &= m_z \bv{\hat{z}} e^{i\beta z},& \quad
\bv{m}_x &= m_x \bv{\hat{x}} e^{i\beta z},& \quad
\bv{m}_y &= m_y \bv{\hat{y}} e^{i\beta z}.
\end{aligned}
\label{eq:linedipoles}
\end{equation}
When given harmonic $e^{-i\omega t}$ time variation, the time derivatives of \eqref{eq:linedipoles} define line currents which produce radiation patterns identical to dipolar scattered fields of the cylinder.\autocite{strickland2015dynamic} We define $\bar{\bar{\alpha}}(\omega, \beta)$ from \eqref{eq:alpha} in terms of the six complex amplitudes of \eqref{eq:linedipoles}, with $\bv{p} = [p_x;\ p_y;\ p_z]$ and $\bv{m} = [m_x;\ m_y;\ m_z]$. Unlike point dipoles, these line dipoles are defined per unit length. To harmonize units among the four quadrants of $\bar{\bar{\alpha}}$, we further rescale the quantities throughout this paper, which are related to the original SI quantities by $\bv{H} = Z\bv{H}^\textrm{SI}$, $\epsilon\bv{p} = \bv{p}^\textrm{SI}/l$, and $\bv{m} = Z\bv{m}^\textrm{SI}/l$, where $\epsilon$ and $Z$ are the permittivity and impedance of the surrounding medium and $l$ is the unit length. Thus, all elements of $\bar{\bar{\alpha}}$ have units of meters squared.

Symmetry restricts the components of $[\bv{E};\ \bv{H}]$ that may couple to $[\bv{p};\ \bv{m}]$, and thus the non-zero components of $\bar{\bar{\alpha}}$. Furthermore, symmetry arguments alone can replicate entirely the structure of $\bar{\bar{\alpha}}$ derived by Strickland et al., thereby justifying the existence of magnetoelectric coupling terms. This ensues by demanding that $\bar{\bar{\alpha}}$ be invariant under all symmetry operations which leave the cylinder invariant. For the cylinder the group of symmetry operations is $D_{\infty h}$, generated by three operations sharing a common point along the cylinder axis: rotation of any angle $\phi$ about the axis, reflection through any plane coplanar with the axis, and inversion about a point. This invariance is established in part by noting that an experiment is unchanged when a rigid transformation is applied to both the scatterer and incident fields, even if the transformation does not correspond to a symmetry operation of the scatterer.\autocite{barron2004molecular} The same $\bar{\bar{\alpha}}$ still predicts the new induced dipoles moments, but only if these can be expressed in terms of the basis set that defines $\bar{\bar{\alpha}}$. The full mathematical details of this argument appear in Appendix \ref{sec:grouptheory}.

We now demonstrate the effects of these $D_{\infty h}$ symmetry operations on the basis set of line dipoles \eqref{eq:linedipoles}. Consider first a rotation of angle $\phi$, $\hat{C}_\infty^\phi$, on the component $\bv{p}_x$. When both incident fields and scatterer are rotated, the dipole induced $\bv{p}_x'$ is similarly rotated,
\begin{equation}
\bv{p}_x' = \hat{C}_\infty^\phi \bv{p}_x = p_x \bv{\hat{x}}' e^{i\beta z'} = p_x(\cos\phi\bv{\hat{x}} + \sin\phi\bv{\hat{y}})e^{i\beta z},
\end{equation}
with primes denoting transformed coordinates. A linear combination of \eqref{eq:linedipoles} is generated, so the new moment can also be represented by the same basis set. But inversion, $\hat{P}$, generates a line dipole outside the set,
\begin{equation}
\bv{p}_x' = \hat{P} \bv{p}_x = p_x \bv{\hat{x}}' e^{i\beta z'} = -p_x \bv{\hat{x}} e^{-i\beta z},
\label{eq:inversion}
\end{equation}
with opposing propagation constant, $-\beta$, so the basis set in which $\bar{\bar{\alpha}}$ is defined is unable to model the new moment.

Thus, invariance under $D_{\infty h}$ is impossible because the basis set \eqref{eq:linedipoles} is not closed under inversion. The basis does not preserve all the symmetries of the cylinder due to the imposed longitudinal $\beta$-variation. In more abstract terms, $\hat{P}\bv{p}_x$ must generate itself or its negative, corresponding to eigenvalues $\pm 1$, since $\hat{P}^2$ is the identity operation.\autocite{barron2004molecular} Both magnetic and electric point dipoles fulfill this requirement, but not the line dipoles \eqref{eq:linedipoles} so these do not transform as an irreducible representation of $D_{\infty h}$.\autocite{hamermesh1962group}

With inversion lacking, $\bar{\bar{\alpha}}$ transforms only under the symmetry operations of point group $C_{\infty v}$, which determines its allowable form,\autocite{barron2004molecular}
\begin{equation}
\begin{bmatrix} p_x\\ p_y \\ p_z \\ m_x \\ m_y \\ m_z \end{bmatrix} = 
\begin{bmatrix}
\alpha^{ee}_{\perp} & 0 & 0 & 0 & -\alpha^{em} & 0\\
0 & \alpha^{ee}_{\perp} & 0 & \alpha^{em} & 0 & 0\\
0 & 0 & \alpha^{ee}_z & 0 & 0 & 0\\
	0 & \alpha^{me} & 0 & \alpha^{mm}_{\perp} & 0 & 0\\
-\alpha^{me} & 0 & 0 & 0 & \alpha^{mm}_{\perp} & 0\\
0 & 0 & 0 & 0 & 0 & \alpha^{mm}_z
\end{bmatrix}
\begin{bmatrix} E_x \\ E_y \\ E_z \\ H_x \\ H_y \\ H_z \end{bmatrix}.
\label{eq:indtensor}
\end{equation}
This is precisely the form obtained by Strickland et al., with the exception of an additional symmetry, $\alpha^{em} = \alpha^{me}$, provided by Onsager relations.\autocite{kong1972theorems,sersic2011magnetoelectric} The straightforward derivation of \eqref{eq:indtensor} considering only $C_{\infty v}$ symmetry is detailed in Appendix \ref{sec:grouptheory}, as is the additional restriction under $D_{\infty h}$, requiring the magnetoelectric coupling to disappear.

To highlight the contrasting symmetry requirements of point and line dipoles, we apply the analysis to a prolate spheroid. We may consider two different spheroids, one which resembles a sphere and one which is elongated to resemble a finite cylinder. Assuming the short axis of the two spheroids are sufficiently subwavelength, the scattered fields may be modeled as either a point dipole or line dipole source. Thus the length of the spheroid determines which of the two symmetry restrictions applies. For the short spheroid, the inversion symmetric $D_{\infty h}$ point group applies and magneto-electric coupling is forbidden. Although the long spheroid possesses the same structural symmetry, the line dipole which models it only has $C_{\infty v}$ symmetry, so magneto-electric coupling appears. Such a size dependence has been experimentally observed, with long wires exhibiting spatial dispersion but not short wires.\autocite{tuniz2012spatial}

\section{TM/TE Formulation}
\label{sec:trans}
Critical to the foregoing discussion is that $\bar{\bar{\alpha}}$ inherits the symmetry properties of its defining line dipoles. Specifically, Strickland et al.\ retrieved $\bar{\bar{\alpha}}$ considering only $e^{i\beta z}$ variation. If instead $\bar{\bar{\alpha}}$ is retrieved for both $e^{\pm i\beta z}$ variations, then magnetoelectric coupling vanishes. At first glance, this alternative formulation appears cumbersome and disadvantageous: two tensors are now necessary to characterize the polarizability, and treating both opposing incidence angles would seem to obscure any connection to spatial dispersion arising from the phase delay of a single plane wave along the cylinder.

On the contrary, we show that this enables a familiar and natural reinterpretation of magnetoelectric coupling, decomposing $\bar{\bar{\alpha}}$ from \eqref{eq:alpha} into 
\begin{equation}
\begin{bmatrix} \bv{p} \\ \bv{m} \end{bmatrix} =
\bar{\bar{\alpha}}^\textrm{TM}
\begin{bmatrix} \bv{E}^\textrm{inc,TM}  \\ \bv{H}^\textrm{inc,TM} \end{bmatrix} +
\bar{\bar{\alpha}}^\textrm{TE}
\begin{bmatrix} \bv{E}^\textrm{inc,TE}  \\ \bv{H}^\textrm{inc,TE} \end{bmatrix},
\label{eq:splitalpha}
\end{equation}
where $\bar{\bar{\alpha}}^\textrm{TM}$ and $\bar{\bar{\alpha}}^\textrm{TE}$ apply only to the TM or TE components of the incident field. Both tensors are diagonal, with
\begin{equation}
\begin{aligned}
\bar{\bar{\alpha}}^\textrm{TM} &=
\diag(\alpha^{ee, \textrm{TM}}_\perp, \alpha^{ee, \textrm{TM}}_\perp, \alpha^{ee, \textrm{TM}}_{z}, \alpha^{mm, \textrm{TM}}_\perp, \alpha^{mm, \textrm{TM}}_\perp, 0),\\
\bar{\bar{\alpha}}^\textrm{TE} &=
\diag(\alpha^{ee, \textrm{TE}}_\perp, \alpha^{ee, \textrm{TE}}_\perp, 0, \alpha^{mm, \textrm{TE}}_\perp, \alpha^{mm, \textrm{TE}}_\perp, \alpha^{mm, \textrm{TE}}_{z}).
\end{aligned}
\label{eq:alphaTMTE}
\end{equation}
Properties of \eqref{eq:splitalpha} seem to follow immediately from the cylindrical geometry. Decomposition of the response into TM and TE incidence is common among structures with infinite translational symmetry, like the Fresnel coefficients of planar interfaces. Meanwhile, inversion symmetry now enforces diagonal responses, as initially expected. More rigorously, these properties follow from the symmetry considerations detailed in Appendix \ref{sec:grouptheory}.

We now consider the transformation between the magnetoelectric formulation of \eqref{eq:alpha} and \eqref{eq:indtensor}, and the TM/TE reformulation \eqref{eq:splitalpha}--\eqref{eq:alphaTMTE}, which is key to the utility of \eqref{eq:splitalpha} in interpreting the magnetoelectric coupling properties of \eqref{eq:alpha}. The two tensors in \eqref{eq:splitalpha} together occupy a vector space twice the size of \eqref{eq:alpha}, seemingly implying that \eqref{eq:splitalpha} requires independent knowledge of both TM and TE components of the incident field. But this is unnecessary, as this knowledge can be deduced from the fields themselves. Indeed, this crucial property forms the foundation of the transformation, and even the reinterpretation of magnetoelectric coupling discussed in Section \ref{sec:discuss}. We exploit the redundancy among the six incident field components of \eqref{eq:alpha}, which are instead specified by only three variables for a given $k$ and polar incidence angle $\theta$: the complex amplitudes of the TM and TE waves and azimuthal incidence angle $\phi$. Thus, $\bar{\bar{\alpha}}^\textrm{TM}$ and $\bar{\bar{\alpha}}^\textrm{TE}$ can be combined into a single polarization independent tensor $\bar{\bar{\alpha}}$. This inverse transformation, from \eqref{eq:splitalpha} to \eqref{eq:alpha}, is more physically illuminating, and is derived below.

For propagation constant $\bv{k} = \bv{\hat{x}}\cos\phi\sin\theta + \bv{\hat{y}}\sin\phi\sin\theta + \bv{\hat{z}}\cos\theta$, the TM and TE plane waves are
\begin{equation}
\begin{aligned}
\bv{E} &= E^\textrm{TM}_0 (-\bv{\hat{x}}\cos\phi\cos\theta - \bv{\hat{y}}\sin\phi\cos\theta + \bv{\hat{z}}\sin\theta) e^{i\bv{k}\cdot\bv{r}},\\
\bv{H} &= E^\textrm{TM}_0 (\bv{\hat{x}}\sin\phi - \bv{\hat{y}}\cos\phi) e^{i\bv{k}\cdot\bv{r}},
\end{aligned}
\label{eq:TMfield}
\end{equation}
and
\begin{equation}
\begin{aligned}
\bv{E} &= E^\textrm{TE}_0 (-\bv{\hat{x}}\sin\phi + \bv{\hat{y}}\cos\phi) e^{i\bv{k}\cdot\bv{r}},\\
\bv{H} &= E^\textrm{TE}_0 (-\bv{\hat{x}}\cos\phi\cos\theta - \bv{\hat{y}}\sin\phi\cos\theta + \bv{\hat{z}}\sin\theta) e^{i\bv{k}\cdot\bv{r}}.
\end{aligned}
\label{eq:TEfield}
\end{equation}

Two crucial ratios between field components, $E_x/H_y$ and $E_y/H_x$, characterize the polarization independently of $\phi$. For example, $E_x/H_y = \cos\theta$ for TM incidence, but equals $1/\cos\theta$ for TE incidence. The $E_x$ and $H_y$ fields resulting from a predetermined superposition of \eqref{eq:TMfield} and \eqref{eq:TEfield} is
\begin{equation}
\begin{bmatrix} E_x \\ H_y\end{bmatrix} =
-\begin{bmatrix} \cos\theta\cos\phi & \sin\phi \\ \cos\phi & \cos\theta\sin\phi\end{bmatrix}
\begin{bmatrix} E^\textrm{TM}_0 \\ E^\textrm{TE}_0 \end{bmatrix}.
\end{equation}
This relation can be inverted to deduce the complex amplitudes $E^\textrm{TM}_0$ and $E^\textrm{TE}_0$ from a given incidence field, thereby decomposing the fields into their TM and TE components,
\begin{equation}
\begin{bmatrix} E_x^\textrm{TM} \\ H_y^\textrm{TM} \\ E_x^\textrm{TE} \\ H_y^\textrm{TE}
\end{bmatrix} =
\frac{1}{\sin^2\theta}
\begin{bmatrix} -\cos^2\theta & \cos\theta \\ -\cos\theta & 1 \\ 1 & -\cos\theta \\ \cos\theta & -\cos^2\theta \end{bmatrix}
\begin{bmatrix} E_x \\ H_y\end{bmatrix},
\label{eq:decomp}
\end{equation}
as denoted by superscripts TM and TE. The fields $E_x$ and $H_y$ have been successfully decomposed without prior knowledge of $E^\textrm{TM}_0$ and $E^\textrm{TE}_0$ independently of azimuthal angle $\phi$. However, a coupling between magnetic and electric fields is introduced, arising without any reference to polarizability tensors.

With $E_x$ and $H_y$ appropriately partitioned, tensors elements from $\bar{\bar{\alpha}}^\textrm{TM}$ and $\bar{\bar{\alpha}}^\textrm{TE}$ in \eqref{eq:alphaTMTE} can be individually applied and their dipole moments subsequently combined to yield
\begin{equation}
\begin{bmatrix} p_x \\ m_y \end{bmatrix} = 
\frac{1}{\sin^2\theta}\begin{bmatrix}
\alpha^{ee,\textrm{TE}}_{\perp} - \alpha^{ee,\textrm{TM}}_{\perp}\cos^2\theta & \cos\theta(\alpha^{ee,\textrm{TM}}_{\perp} - \alpha^{ee,\textrm{TE}}_{\perp})\\
\cos\theta(\alpha^{mm,\textrm{TE}}_{\perp} - \alpha^{mm,\textrm{TM}}_{\perp}) & \alpha^{mm,\textrm{TM}}_{\perp} - \alpha^{mm,\textrm{TE}}_{\perp}\cos^2\theta
\end{bmatrix}
\begin{bmatrix} E_x \\ H_y \end{bmatrix},
\label{eq:EyHxtensor}
\end{equation} 
where off-diagonal elements can be directly retraced to the coupling present in \eqref{eq:decomp}.  Repeating the procedure for $E_y$ and $H_x$ produces the remaining off-diagonal elements, combining $\bar{\bar{\alpha}}^\textrm{TM}$ and $\bar{\bar{\alpha}}^\textrm{TE}$ into $\bar{\bar{\alpha}}$ in the form of \eqref{eq:indtensor}, where
\begin{gather}
\begin{aligned}
\alpha^{ee}_\perp &= k_\perp^{-2}(k^2\alpha^{ee,\textrm{TE}}_\perp - \beta^2\alpha^{ee,\textrm{TM}}_\perp), & \quad
\alpha^{mm}_\perp &= k_\perp^{-2}(k^2\alpha^{mm,\textrm{TM}}_\perp - \beta^2\alpha^{mm,\textrm{TE}}_\perp),
\label{eq:transform}
\end{aligned}\\
k_\perp^{-2}\beta k(\alpha^{ee,\textrm{TE}}_\perp - \alpha^{ee,\textrm{TM}}_\perp) = \alpha^{em} = 
\alpha^{me} = k_\perp^{-2}\beta k(\alpha^{mm,\textrm{TM}}_\perp - \alpha^{mm,\textrm{TE}}_\perp),
\label{eq:transformem}
\end{gather} 
and $k_\perp^2 = k^2 - \beta^2$, thus defining the inverse transformation from \eqref{eq:splitalpha} to \eqref{eq:alpha}. In \eqref{eq:transformem}, the second equality follows from Onsager relations.\autocite{kong1972theorems,sersic2011magnetoelectric} The axial components of the two formulations are identical and unaffected by the transformation,
\begin{align}
\alpha_z^{ee} &= \alpha_z^{ee,\textrm{TM}}, & \alpha_z^{mm} &= \alpha_z^{mm,\textrm{TE}},
\label{eq:axialeq}
\end{align}
since axial fields $E_z$ and $H_z$ are already exclusively associated with TM or TE incidence, respectively.

The forward transformation, from \eqref{eq:alpha} to \eqref{eq:splitalpha}, can be obtained by inverting the system of equations \eqref{eq:transform}--\eqref{eq:transformem}, hence establishing the mathematical equivalence between the magnetoelectric formulation and its TM/TE reformulation. Alternatively, the transformation can be derived directly from \eqref{eq:alpha} and \eqref{eq:indtensor} if exclusively TM or TE fields, \eqref{eq:TMfield}--\eqref{eq:TEfield}, are used as inputs. Then by exploiting the characteristic ratios $E_x/H_y$ and $E_y/H_x$, \eqref{eq:indtensor} can always be shown to produce the same mathematical result as a diagonal tensor, of the form \eqref{eq:alphaTMTE}. Furthermore, \eqref{eq:splitalpha}--\eqref{eq:alphaTMTE} can be derived $\emph{ab initio}$, and its involved details will be supplied in a forthcoming paper. Note that this transformation is not the diagonalization of $\bar{\bar{\alpha}}$, in part because it produces two tensors \eqref{eq:alphaTMTE} that together occupy a vector space twice the size of \eqref{eq:indtensor}. 

When applied to the polarizability tensor derived by Strickland et al., the transformation produces explicit expressions for tensor elements in \eqref{eq:splitalpha}--\eqref{eq:alphaTMTE},
\begin{equation}
\begin{aligned}
\alpha_\perp^{ee,\textrm{TM}} &= -8ik_\perp^{-2}\left(-c_1^\textrm{TM} + \frac{k}{\beta}c_1^x\right), & \quad
\alpha_\perp^{mm,\textrm{TM}} &= -8ik_\perp^{-2}\left(c_1^\textrm{TM} - \frac{\beta}{k}c_1^x\right),\\
\alpha_\perp^{ee,\textrm{TE}} &= -8ik_\perp^{-2}\left(c_1^\textrm{TE} - \frac{\beta}{k}c_1^x\right), & \quad
\alpha_\perp^{mm,\textrm{TE}} &= -8ik_\perp^{-2}\left(-c_1^\textrm{TE} + \frac{k}{\beta}c_1^x\right),
\end{aligned}
\label{eq:strickpol}
\end{equation}
where $c_1^\textrm{TM}$ and $c_1^\textrm{TE}$ are the TM and TE Mie scattering coefficients for dipolar cylindrical harmonic incidence fields, and $c_1^x$ is the cross-coupling between TM and TE during scattering. These are defined according to the notation of Strickland et al.,\autocite{strickland2015dynamic} which feature minor differences relative to common textbook definitions.\autocite{bohren1983absorption}

The explicit form of \eqref{eq:strickpol} shows that the transformation preserves the units of $\bar{\bar{\alpha}}$, the common unit being meters squared. Note also that $\bar{\bar{\alpha}}^\textrm{TM}(\omega, \beta)$ and $\bar{\bar{\alpha}}^\textrm{TE}(\omega, \beta)$ both depend on $\beta/k$, corresponding to $\cos\theta$. Furthermore, each tensor element implicitly depends on $\theta$ via the Mie coefficients. Angle dependent tensors, though atypical, do not impede the subsequent application of homogenization theories. The resulting constitutive relations also depend on the angle of incidence.\autocite{silveirinha2006nonlocal} The angle dependence arises due to the longitudinal invariance of the cylinders, which discriminates all cylinder properties according to the harmonic variation $\beta$. However, angle dependent polarizabilities are not exclusive to cylinders, and arise whenever the gradient of the fields are important.\autocite{varault2013multipolar,grahn2013multipole}

\section{Discussion}
\label{sec:discuss}
Using the systematic and universal language of group representation theory, we have demonstrated that the magnetoelectric coupling of infinite cylinders derived by Strickland et al.\ is entirely consistent with the existing literature on symmetry and magnetism. Magnetoelectric coupling of microscopic polarizability $\bar{\bar{\alpha}}$ only vanishes when both the scatterer and the basis of dipoles representing it are inversion symmetric. The bulk of the literature concerns point dipoles or their homogenization,\autocite{raab2004multipole,barron2004molecular,landau1984electrodynamics,birss1964symmetry,odell1970electrodynamics,marques2002role,padilla2007group,baena2007systematic} which are closed under inversion, so inversion symmetric scatterers automatically have zero magnetoelectric coupling.

Meanwhile, no unique choice exists for defining line dipoles. Using $e^{i\beta z}$ longitudinal variation as in \eqref{eq:linedipoles}, seemingly the most natural choice as it matches the spatial variation of obliquely incident plane waves, has the disadvantage that the resulting dipoles do not retain all the symmetries of the cylinders which they model. Since polarizability $\bar{\bar{\alpha}}$ is defined in terms of these line dipoles, this introduces a $\beta$-dependence to $\bar{\bar{\alpha}}(\omega, \beta)$, whose symmetry properties may differ from polarizability tensors that are functions of $\omega$ alone, $\bar{\bar{\alpha}}(\omega)$. The consequent non-zero magnetoelectric coupling may be said to arise from oblique incidence, which breaks the cylinder's inversion symmetry.\autocite{strickland2015dynamic} However, the analysis reveals a subtle distinction compared to the symmetry breaking of extrinsic chirality, which arises even for point dipoles due to the restricted orientation of the incidence relative to the scatterer.\autocite{saadoun1992reciprocal,tretyakov1993eigenwaves,barron2004molecular,efrati2014orientation} 

Magnetoelectric coupling is not a necessary consequence of a polarizability tensor that has $\beta$-dependence, and alternatively tensors that are valid for both $e^{\pm i\beta z}$ incidence can be retrieved. Two tensors \eqref{eq:splitalpha}--\eqref{eq:alphaTMTE} result, both of which are diagonal, thus conforming to symmetry expectations of the cylindrical geometry. Far from being arbitrary, the response partitions into its TM and TE contributions, $\bar{\bar{\alpha}}^\textrm{TM}$ and $\bar{\bar{\alpha}}^\textrm{TE}$. The TM/TE reformulation predicts identical results to the magnetoelectric formulation $\bar{\bar{\alpha}}$ given by \eqref{eq:alpha} and \eqref{eq:indtensor}, their equivalence established by the transformation \eqref{eq:transform}--\eqref{eq:transformem}. 

The two formulations can thus be regarded as complementary, and their practical utility depends on the application. The magnetoelectric form \eqref{eq:indtensor} applies to general incidence, while the simpler diagonal forms \eqref{eq:alphaTMTE} apply only to restricted incidence, though we show that decomposition of arbitrary incidence fields into its TM and TE components can always be achieved. Correspondingly, homogenization based on the magnetoelectric form produces more generally applicable constitutive relations, but the resulting expressions are more complex. In the electrodynamic limit, this typically proceeds by accounting for the multiple scattering from all cylinders in the lattice, then averaging fields over the unit cell. While the extension of the former to magnetoelectric tensors is relatively straightforward,\autocite{lunnemann2013optical} simplifications result from the latter if $\bar{\bar{\alpha}}$ is diagonal. For example, field averaging can be achieved by taking the matrix inverse of $\bar{\bar{\alpha}}$,\autocite{silveirinha2006nonlocal} but a more complex dependence on magnetoelectric coupling arises even if a scalar magnetoelectric term is present.\autocite{alu2011first} Thus, the diagonal formulation is preferable and applicable if the response of the opposing polarization can be neglected. This occurs for example at frequencies where the difference between TM and TE polarizabilities is small, and where cross-polarization scattering is small or zero, such as for perfectly conducting cylinders or dilute arrays of weak scatters.

The second primary benefit of the TM/TE reformulation is the insights it provides, discussed below and in Sections \ref{sec:numerics} and \ref{sec:symmetry}. With the transformation in hand, we may set aside mathematically-oriented group theory arguments for the appearance of magnetoelectric coupling, and henceforth seek physically motivated explanations into its origins and properties. The TM/TE reformulation is not only simpler by virtue of its diagonal responses, but also coincides with the familiar separation of cylinder properties into TM and TE components. Note that the following observations apply to the in-plane elements of $\bar{\bar{\alpha}}$, $\bar{\bar{\alpha}}^\textrm{TM}$, and $\bar{\bar{\alpha}}^\textrm{TE}$, since the axial polarizabilities are identical between the two formulations, as per \eqref{eq:axialeq}.

The magnetoelectric coupling of $\bar{\bar{\alpha}}$ can be conceptualized as accounting for the difference between TM and TE polarizabilities, combining two differing diagonal responses, $\bar{\bar{\alpha}}^\textrm{TM}$ and $\bar{\bar{\alpha}}^\textrm{TE}$, into a single tensor with off-diagonal terms. This equivalence stems from the freedom to attribute the induced dipoles to either $\bv{E}$ or $\bv{H}$ fields, as these are indivisible components of plane wave inputs \eqref{eq:TMfield}--\eqref{eq:TEfield}. From \eqref{eq:transform}, we arrive at the interpretation that the diagonal terms $\alpha_\perp^{ee}$ and $\alpha_\perp^{mm}$ within the magnetoelectric formulation $\bar{\bar{\alpha}}$ of \eqref{eq:indtensor} correspond to a weighted average of the TM and TE polarizabilities of \eqref{eq:alphaTMTE}. Then, the off-diagonal magnetoelectric coupling terms $\alpha^{em}$ and $\alpha^{me}$ encode the difference \eqref{eq:transformem} between TM and TE responses. Note from \eqref{eq:strickpol} that this difference only requires contrast between the TM and TE scattering coefficients, $c_1^\textrm{TM}$ and $c_1^\textrm{TE}$. A non-zero cross-coupling coefficient $c_1^x$ is unnecessary, a case we revisit in the next section.

The structure of the non-zero off-diagonal terms is also revealing, corresponding precisely to the characteristic ratios $E_x/H_y$ and $E_y/H_x$ that enable an arbitrary incidence to be decomposed into its TM and TE components, via \eqref{eq:decomp}. This decomposition is achieved without foreknowledge of incidence polarization, but introduces a coupling between $\bv{E}$ and $\bv{H}$ fields which originates entirely from the geometry of plane waves. For example, both $E_x$ and $H_y$ contribute to $E_x^\textrm{TM}$, the TM component of $E_x$, with strengths that depends implicitly on incidence polarization via the ratio $E_x/H_y$. The transformation reveals that the off-diagonal magnetoelectric coupling terms emerge from this decomposition, which is embedded within the polarization tensor $\bar{\bar{\alpha}}$ to ensure that the correct induced moments $\bv{p}$ and $\bv{m}$ are predicted regardless of polarization. Thus, the magnetoelectric contribution to the total $\bv{p}$ and $\bv{m}$ also depends implicitly on incidence polarization via the incidence fields. We may conclude that the magnetoelectric and TM/TE formulations are fully interchangeable, and their sole distinction is whether this decomposition has been incorporated into the tensor.

\subsection{Quantitative Properties and Numerical Examples}
\label{sec:numerics}

Symmetry and spatial dispersion arguments concern the existence of magnetoelectric coupling,\autocite{fietz2010current} but are silent on its quantitative behavior. We now exploit the interchangeability of the two formulations to gain such insight. Key is \eqref{eq:transformem}, tracing the magnitude of $\alpha^{em}$ to the difference between polarizations in both the electric and magnetic polarizabilities, which are given explicitly in \eqref{eq:strickpol}. We examine both dielectric and metallic cylinders, discussing both the scaling of magnetoelectric coupling at long wavelengths and its strength at resonance. 

\begin{figure}[tb]
\begin{center}
\subfloat{\includegraphics{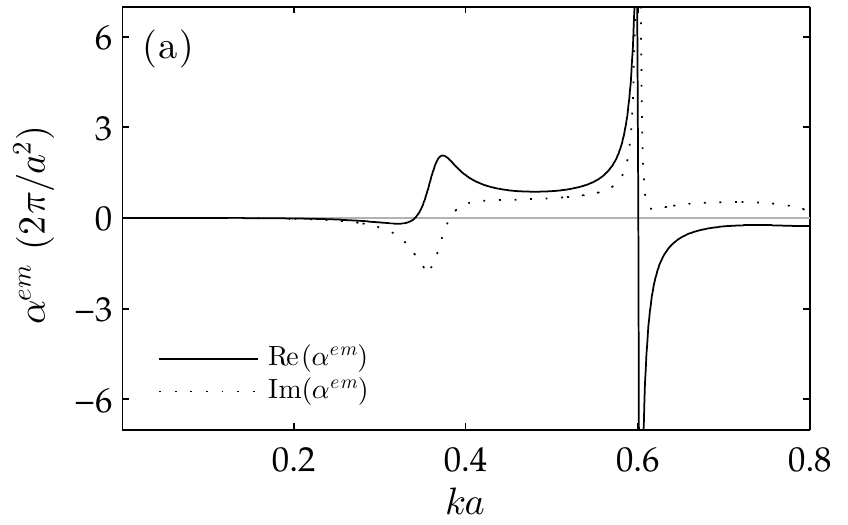}}\\
\noindent\makebox[\textwidth]{%
\subfloat{\includegraphics{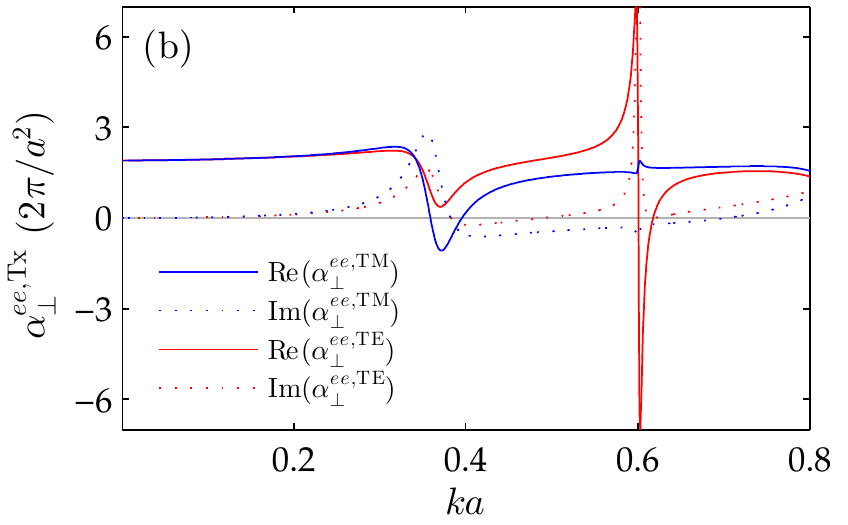}}
\subfloat{\includegraphics{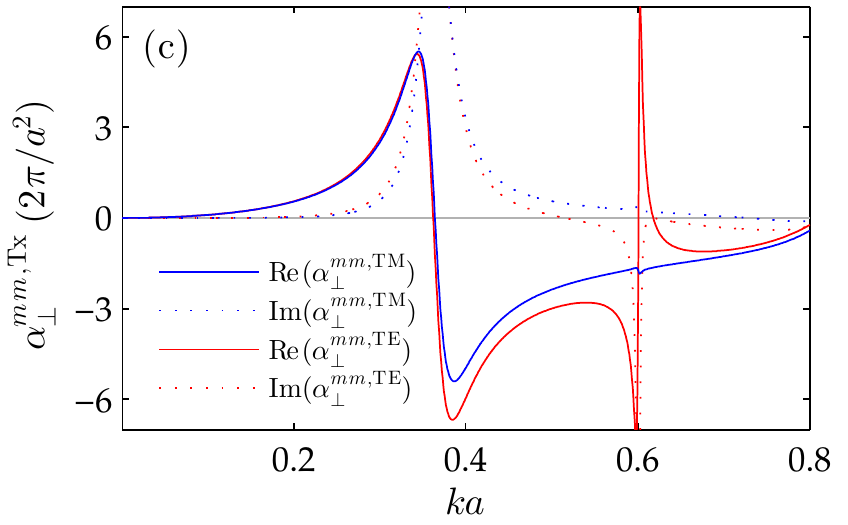}}
}
\caption[]{Various components of polarizability, indicated by the legends, as a function of wavenumber $k$ and cylinder radius $a$. The geometry is a dielectric cylinder $\epsilon = 40$ in vacuum excited at polar angle of incidence $\theta = \pi/4$. Shown are (a) magnetoelectric coupling, (b) electric polarizabilities, and (c) magnetic polarizabilities, where (a) is calculated using \eqref{eq:transformem} and (b)--(c) using \eqref{eq:strickpol}. TM and TE polarizabilities are respectively plotted in blue and red, while real and imaginary parts are respectively indicated by solid and dashed lines.}
\label{fig:dielectric}
\end{center}
\end{figure}

We consider first the scaling of a high index dielectric cylinder in vacuum, using parameters similar to the examples of Strickland et al. Plotted in Fig.\ \ref{fig:dielectric}(a) is $\alpha^{em}$ at oblique incidence, showing its insignificant magnitude at long wavelengths due to its weak $k^4$ scaling. Such scaling can be anticipated neither from Onsager relation requirements,\autocite{sersic2011magnetoelectric,sersic2012ubiquity} nor from arguments based either on group theory or the spatial dispersion of magnetoelectric coupling.\autocite{li2007non,fietz2010current,alu2011restoring} But it may be deduced by considering the magnetic polarizabilities $\alpha_\perp^{mm,\textrm{TM}}$ and $\alpha_\perp^{mm,\textrm{TE}}$ individually. As seen in Fig.\ \ref{fig:dielectric}(c), these both have identical quadratic scaling, so their difference is quartic to lowest order. The same quartic dependence of $\alpha^{em}$ can also be deduced from electric polarizabilities $\alpha_\perp^{ee,\textrm{TE}}$ and $\alpha_\perp^{ee,\textrm{TM}}$ in Fig.\ \ref{fig:dielectric}(b), due to the symmetry between $\alpha^{em}$ and $\alpha^{me}$ guaranteed by the Onsager relations embedded in \eqref{eq:transformem}.

As already observed by Strickland et al., magnetoelectric coupling is non-zero even for perfectly conducting cylinders. This finding is unexpected if one adopts the seemingly plausible, but ultimately erroneous, view that magnetoelectric coupling originates from the polarization cross-coupling between TM and TE waves during scattering at oblique incidence. The perfectly conducting case violates this view, as $c_1^x$ from \eqref{eq:strickpol} is zero for all incidence angles.\autocite{strickland2015dynamic,ruppin2006scattering} This is exemplified by the stronger scaling of $\alpha^{em}$ at long wavelengths, approximately proportional to $k^2$. Unlike the dielectric case, both TM and TE polarizabilities now converge to a constant at $k=0$, given by $\alpha_\perp^{ee,\textrm{TM}} = \alpha_\perp^{ee,\textrm{TE}} = -\alpha_\perp^{mm,\textrm{TM}} = -\alpha_\perp^{mm,\textrm{TE}} = 2\pi a^2$. But they differ at the quadratic term, so $\alpha^{em} \approx k^2a^2\cos\theta(-2\log(ka\sin\theta)+i\pi)\pi a^2$. Again, the origin of this stronger scaling is inaccessible to group theory or spatial dispersion arguments. We note in passing that axial polarizability $\alpha_z^{ee}$ of perfect conductors tends towards infinity at long wavelengths, corresponding to the strong axial polarizability of wire grid polarizers.

\begin{figure}[tb]
\begin{center}
\noindent\makebox[\textwidth]{%
\subfloat{\includegraphics{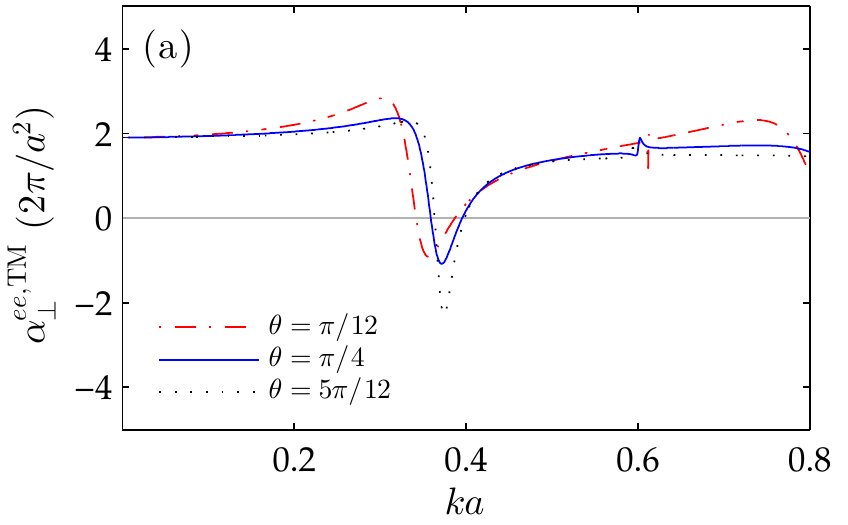}}
\subfloat{\includegraphics{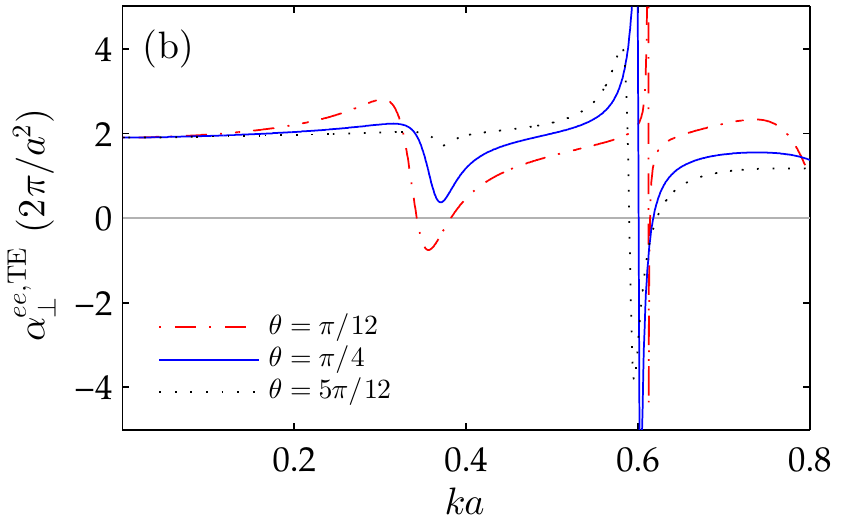}}
}
\caption[]{As in Fig.\ \ref{fig:dielectric}(b) but shows TM and TE polarizabilities at differing polar incidence angles $\theta = \pi/12$ (solid blue), $\theta = \pi/4$ (dash-dotted red), and $\theta = 5\pi/12$ (dotted black). At near-normal incidence, the first resonance manifests primarily in the TE response and the second resonance in the TM. At oblique incidence this continues to be true, though to a lesser extent for the first resonance.}
\label{fig:angle}
\end{center}
\end{figure}

Moving away from the long wavelength limit, the magnetoelectric coupling of dielectric cylinders becomes prominent at resonance. Here, TM and TE responses may differ substantially, e.g., at the second resonance near $ka = 0.6$ in Fig.\ \ref{fig:dielectric}. The peak present for the TE response is absent for TM. Such contrast is likely a general feature at resonance as evidenced by plotting polarizabilities as a function of polar incidence angle $\theta$, as in Fig.\ \ref{fig:angle}. Each resonance is associated with either TM or TE incidence at near-normal incidence, and even at oblique incidence the resonances often retain their TM or TE character. This permits some intuition into magnetoelectric behavior at oblique incidence to be deduced from the normal incidence case, where a clear demarcation between TM or TE resonances exists. A second consequence of \eqref{eq:transformem} is that magnetoelectric coupling resonances do not occur independently of electric and magnetic resonances. This is most apparent at the second resonance of Fig.\ \ref{fig:dielectric}, where only TE polarizabilities show a response. Then, the full set of equalities in \eqref{eq:transformem} implies similarity between the magnitudes of electric, magnetic, and magnetoelectric resonances. This contrasts with scatterers that lack inversion symmetry, which allow magnetoelectric coupling to differ in magnitude from either the electric or magnetic resonance.\autocite{alaee2015magnetoelectric}

So far, the discussion has focused primarily on the real part of polarizability, but we now specifically address the imaginary parts of Figs \ref{fig:dielectric}(b)-(c). Both positive and negative imaginary parts are observed even for lossless cylinders. This does not violate any energy conservation requirements, as we can verify that the energy extinguished from the incident field matches the energy radiated by $\bv{p}$ and $\bv{m}$. Appendix \ref{sec:sipekran} derives the Sipe-Kranendonk relations \eqref{eq:skeez}--\eqref{eq:skte}, which express this energy balance directly in terms of tensor elements within $\bar{\bar{\alpha}}^\textrm{TM}$ or $\bar{\bar{\alpha}}^\textrm{TE}$. Specifically, \eqref{eq:sktm} and \eqref{eq:skte} state that the imaginary parts of both electric and magnetic polarizabilities must be considered together to satisfy energy conservation.

\subsection{Symmetry Properties}
\label{sec:symmetry}

\begin{figure}[tb]
\begin{center}
\includegraphics[width=\textwidth]{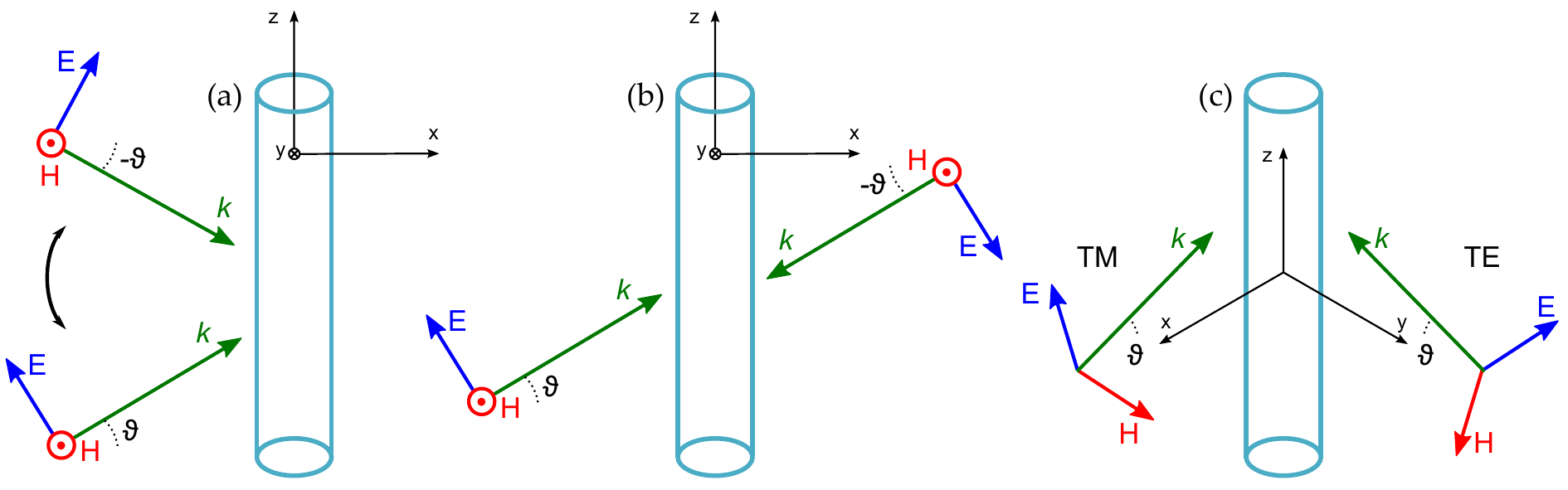}
\caption[]{Three examples of incident plane waves demonstrating various properties of the magnetoelectric coupling, where $\vartheta$ indicates inclination above the $x$-$y$ plane. (a) Changing the angle of incidence $\vartheta \rightarrow -\vartheta$, or equivalently $\beta \rightarrow -\beta$, also changes the sign of the induced moment $p_x$. This originates from the change in sign of $E_x$, while $H_y$ and $E_z$ remain unchanged. (b) Two counterpropagating waves are incident, arranged so that $E_x$ and $E_z$ fields cancel while $H_y$ fields constructively interfere. No electric dipole moments $\bv{p}$ are induced. (c) Switching to isometric view, two incoming plane waves of both TM and TE polarization are shown, with identical $\vartheta$ but orthogonal azimuthal angles $\phi$. Magnitudes are chosen such that $E_x$ fields cancel, but $H_y$ fields do not. A $p_x$ moment is nevertheless induced. The only other non-zero fields are $E_z$ and $H_z$.}
\label{fig:incidence}
\end{center}
\end{figure}
A striking symmetry property of the magnetoelectric coupling, as noted by Strickland et al., is its odd dependence on wavevector. While this property can be derived using mathematical arguments,\autocite{fietz2010current} we show that its simple geometric origin is revealed once the role of the off-diagonal terms as accounting for the difference between TM and TE responses is identified. We discuss a key practical consequence of this odd symmetry, which prevents the direct numerical retrieval of the magnetoelectric coupling.

Odd symmetry, $\bar{\bar{\alpha}}^{em}(\omega, \beta) = -\bar{\bar{\alpha}}^{em}(\omega, -\beta)$, implies that the sign of magnetoelectric coupling depends on the choice of mathematical coordinate axes. This curious feature disappears in the TM/TE formulation, since the elimination of the off-diagonal terms restores the one-to-one correspondence between the electric dipole moments and their true origin, the electric fields. Consider the effect of changing the angle of incidence of an impinging plane wave, $\beta \rightarrow -\beta$, as in Fig.\ \ref{fig:incidence}(a). Field components $E_z$ and $H_y$ remain unchanged, but $E_x$ changes sign. From the diagonal TM tensor $\bar{\bar{\alpha}}^\textrm{TM}$ \eqref{eq:alphaTMTE}, it follows that $p_x$ should also change sign. Under the magnetoelectric view however, the unchanged $H_y$ acts as a proxy for the $E_x$ field and also contributes to $p_x$. Since an identical result must be predicted, $\bar{\bar{\alpha}}^{em}$ is required to change sign to match the sign change in $p_x$. This example illustrates in qualitative terms that spatial dispersion of magnetoelectric coupling within $\bar{\bar{\alpha}}(\omega, \beta)$ is fully captured by the TM/TE decomposition, which in turn originates from the infinite translational symmetry of the cylinder.

Odd symmetry holds consequences for an important incidence configuration consisting of counterpropagating waves shown Fig.\ \ref{fig:incidence}(b). In the literature, such configurations are commonly used alongside numerical methods to retrieve polarizability tensors.\autocite{ishimaru2003generalized,vallecchi2011collective,morits2011isotropic} At sufficiently long wavelengths, $\bv{E}$ fields destructively interfere over the cross-section of the scatterer, leaving only $H_y$ non-zero. In the context of magnetoelectric coupling, any induced electric moments can thus be unambiguously attributed to $H_y$. However, magnetoelectric coupling can never be isolated with this configuration, since both impinging plane waves are of the same polarization. No electric moment will be induced in the absence of $\bv{E}$ fields, which is an immediate consequence of diagonal $\bar{\bar{\alpha}}^\textrm{TM}$, \eqref{eq:alphaTMTE}. 

This property is insignificant for the retrieval of the polarizability tensors of cylinders with circular cross-sections since exact Mie solutions are available. Strickland et al.\ exploited this by matching the scattered Mie fields directly to the radiated fields of line dipoles, bypassing the need to use counterpropagating waves.\autocite{strickland2015dynamic} However, the symmetry analysis of Section \ref{sec:group} can be used to show that other cylinders with inversion symmetry, such as those with square and elliptical cross-sections, also exhibit magnetoelectric coupling. Here, the inability to retrieve the magnetoelectric response using counterpropagating waves becomes problematic. Our transformation \eqref{eq:transform}--\eqref{eq:transformem} provides a viable alternative route, allowing the TM and TE responses of the form \eqref{eq:alphaTMTE} to be independently retrieved, which can be combined into a single tensor of the form \eqref{eq:indtensor} applicable to general incidence.

Only when both TM and TE waves are simultaneously impinging can magnetoelectric coupling be isolated by a superposition of plane waves. In contrast to the previous example, Fig.\ \ref{fig:incidence}(c) shows two plane waves of orthogonal polarizations impinging at orthogonal azimuthal angles $\phi$. The amplitudes are chosen such that in-plane $\bv{E}$ fields cancel but in-plane $\bv{H}$ fields do not. An in-plane electric dipole moment $p_x$ then appears to be induced by the only non-zero in-plane field $H_y$, seemingly a manifestation of magnetoelectric coupling. Alternatively, the induced $p_x$ can be considered an artifact of the difference between TM and TE polarizabilities. Even though $E_x$ is zero, this is achieved by superposing plane waves of different polarizations, so a remnant electric dipole moment $p_x$ is induced equal to the difference between $\alpha_\perp^{ee,\textrm{TM}}$ and $\alpha_\perp^{ee,\textrm{TE}}$, as in \eqref{eq:alphaTMTE} or more explicitly in \eqref{eq:strickpol}. In this analysis, the non-zero axial fields, $E_z$ and $H_z$, are ignored, as no coupling exists between axial and in-plane fields and moments in either formulation. These provide concrete examples of magnetoelectric coupling accounting for the difference between TM and TE polarizabilities and of the general interchangeability between the magnetoelectric and TM/TE interpretations.

\section{Summary}
We have provided two independent explanations for the appearance of magnetoelectric coupling in the microscopic polarizability of an infinite cylinder, which defies initial expectations due to the cylinder's center of inversion. Firstly, we provide formal symmetry arguments using group representation theory, which restricts the allowable non-zero elements of polarizability $\bar{\bar{\alpha}}$ based on the set of symmetry operations of the cylinder's point group, $D_{\infty h}$. Unlike point dipoles, the line dipoles \eqref{eq:linedipoles} used to model the cylinder response are not closed under inversion \eqref{eq:inversion}, due to the imposed $e^{i\beta z}$ variation. The lower symmetry of the line dipoles only transforms as an irreducible representation of $C_{\infty v}$, which then constrains $\bar{\bar{\alpha}}$ to have the form \eqref{eq:indtensor}, with non-zero magnetoelectric coupling. This resolves the discrepancy with existing group-theoretic symmetry restrictions on $\bar{\bar{\alpha}}$.

Secondly, we decompose $\bar{\bar{\alpha}}$ into its TM and TE components \eqref{eq:splitalpha}--\eqref{eq:alphaTMTE}, valid for both $e^{\pm i\beta z}$ incidence and thus eliminating the need for magnetoelectric coupling. A transformation \eqref{eq:transform}--\eqref{eq:transformem} is derived between the magnetoelectric form \eqref{eq:alpha} and \eqref{eq:indtensor} and its TM/TE reformulation, enabled by the ability to decompose an arbitrary incidence into its TM and TE components \eqref{eq:decomp}. The transformation demonstrates that diagonal terms within the magnetoelectric formulation \eqref{eq:indtensor} are the weighted average of TM and TE polarizabilities \eqref{eq:transform}, while off-diagonal magnetoelectric coupling terms account for the difference \eqref{eq:transformem}. The magnetoelectric contribution to the induced dipoles varies implicitly with incidence polarization, thus merging two dissimilar responses \eqref{eq:alphaTMTE} onto a single tensor \eqref{eq:indtensor}. The TM/TE reformulation predicts identical results to the magnetoelectric formulation, so their simpler diagonal forms facilitate simpler homogenization procedures.

The simplicity and familiarity of the TM/TE decomposition enables ready physical insights into the behavior of magnetoelectric coupling. The odd dependence of magnetoelectric coupling on $\beta$ is immediately apparent from the geometry of plane waves as a function of polar incidence angle, as is the inability to isolate magnetoelectric coupling using counterpropagating waves. A key advantage of the TM/TE formulation is the quantitative insights it also offers, via \eqref{eq:transformem}. At long wavelengths, magnetoelectric coupling scales as $k^4$ for dielectrics and as $k^2$ for perfect conductors, which follow from the contrast between TM and TE responses. This contrast becomes pronounced at Mie resonances, so \eqref{eq:transformem} predicts that the magnitude of magnetoelectric coupling becomes comparable to the magnitudes of both electric and magnetic resonances and that these resonances all occur simultaneously.

\section*{Acknowledgments}
The authors would like to thank Andrea Al\`{u} and David J.\ Bergman for helpful discussions. PYC acknowledges fellowship support from the Tel Aviv University Center for Nanoscience and Nanotechnology.

\appendix
\section{Derivation of Symmetry Restrictions}
\label{sec:grouptheory}
This section derives the restrictions on $\bar{\bar{\alpha}}$ from geometric symmetries alone, culminating in the form \eqref{eq:indtensor}. Use of theorems and tools from group representation theory is minimized, though some steps may be simplified or stated more rigorously with their aid. 

Consider the polarizability tensor from \eqref{eq:alpha},
\begin{equation}
\begin{bmatrix} \bv{p} \\ \bv{m} \end{bmatrix} = 
\begin{bmatrix}
\bar{\bar{\alpha}}^{ee} & \bar{\bar{\alpha}}^{em}\\
\bar{\bar{\alpha}}^{me} & \bar{\bar{\alpha}}^{mm}
\end{bmatrix}
\begin{bmatrix} \bv{E} \\ \bv{H} \end{bmatrix},
\label{eq:blockalpha}
\end{equation}
suppressing for brevity the superscripts on $\bv{E}$ and $\bv{H}$. We may simplify the derivation by treating each quadrant of \eqref{eq:blockalpha} separately, beginning with
\begin{equation}
\bv{p} = \bar{\bar{\alpha}}^{ee} \bv{E}.
\label{eq:preop}
\end{equation}
Consider first a rotation of the scatterer. In general, a scatterer is not invariant under rotation, so $\bv{p}$ and $\bv{E}$ are related by a new polarizability tensor,
\begin{equation}
\bv{p} = (\bar{\bar{\alpha}}^{ee})' \bv{E}.
\label{eq:postop}
\end{equation}
If however both scatterer and incidence fields are corotated, then the experiment is unchanged, so induced dipoles are similarly corotated and are predicted by the original polarizability tensor,
\begin{equation}
\bv{p}' = \bar{\bar{\alpha}}^{ee} \bv{E}'.
\end{equation}
This is true regardless of any symmetries of the structure. The primed quantities are related to the originals by the transformation $T$, which represents the rotation,
\begin{align}
\bv{p}' &= T\bv{p}, & \bv{E}' &= T\bv{E}.
\label{eq:T}
\end{align}
Thus,
\begin{equation}
(\bar{\bar{\alpha}}^{ee})' = T\bar{\bar{\alpha}}^{ee}T^{-1}.
\label{eq:symcond1}
\end{equation}
By the invariance of electromagnetism under parity,\autocite{barron2004molecular} \eqref{eq:symcond1} also applies to reflections, or more generally any rigid transformation $T$. The caveat is that \eqref{eq:T} must be true: the set $\bv{p}$ is closed, which enables the new dipole moments $\bv{p}'$ to be expressed as a linear combination of $\bv{p}$ given by $T$. 

If now the scatterer is invariant under transformation $T$, the initial experiment is recreated when the scatterer is rotated even if the incidence is not corotated. This furthermore implies
\begin{equation}
(\bar{\bar{\alpha}}^{ee})' = \bar{\bar{\alpha}}^{ee}.
\label{eq:symcond2}
\end{equation}
The combination of \eqref{eq:symcond1} and \eqref{eq:symcond2} yields the symmetry restrictions. The analysis for the other quadrants of \eqref{eq:blockalpha} follows identically, noting that pseudo-vectorial magnetic quantities transform by $-T$ if the transformation includes a reflection, which can be be deduced from the sign of $\det(T)$.

The complete set of restrictions on $\bar{\bar{\alpha}}$ under the $C_{\infty v}$ group can be derived by considering all possible reflections $\sigma_v(\phi)$, where $\phi$ is the azimuthal angle defining a reflection plane coplanar with the cylinder axis. In matrix form,
\begin{equation}
T = \begin{bmatrix}
\cos 2\phi & \sin 2\phi & 0\\
\sin 2\phi & -\cos 2\phi & 0\\
0 & 0 & 1
\end{bmatrix},
\end{equation}
which operates on 
\begin{equation}
\bar{\bar{\alpha}}^{ee} = \begin{bmatrix}
\alpha^{ee}_{xx} & \alpha^{ee}_{xy} & \alpha^{ee}_{xz} \\ 
\alpha^{ee}_{yx} & \alpha^{ee}_{yy} & \alpha^{ee}_{yz} \\
\alpha^{ee}_{zx} & \alpha^{ee}_{zy} & \alpha^{ee}_{zz} 
\end{bmatrix}.
\end{equation}
By \eqref{eq:symcond1}, this gives
\begin{equation}
\begin{split}
&\begin{bmatrix}
\alpha^{ee}_{xx} \cos^2 2\phi + \alpha^{ee}_{yy} \sin^2 2\phi + \frac{1}{2}(\alpha^{ee}_{xy} + \alpha^{ee}_{yx}) \sin 4\phi &
-\alpha^{ee}_{xy} \cos^2 2\phi + \alpha^{ee}_{yx} \sin^2 2\phi + \frac{1}{2}(\alpha^{ee}_{xx} - \alpha^{ee}_{yy}) \sin 4\phi &
0 \\
-\alpha^{ee}_{yx} \cos^2 2\phi + \alpha^{ee}_{xy} \sin^2 2\phi + \frac{1}{2}(\alpha^{ee}_{xx} - \alpha^{ee}_{yy}) \sin 4\phi &
\alpha^{ee}_{yy} \cos^2 2\phi + \alpha^{ee}_{xx} \sin^2 2\phi + \frac{1}{2}(\alpha^{ee}_{xy} + \alpha^{ee}_{yx}) \sin 4\phi &
0 \\
0 & 0 & 0
\end{bmatrix}\\
+ &\begin{bmatrix}
0 & 0 & \alpha^{ee}_{xz} \cos 2\phi + \alpha^{ee}_{yz} \sin 2\phi \\
0 & 0 & \alpha^{ee}_{xz} \sin 2\phi - \alpha^{ee}_{yz} \cos 2\phi \\
\alpha^{ee}_{zx} \cos 2\phi + \alpha^{ee}_{zy} \sin 2\phi &
\alpha^{ee}_{zx} \sin 2\phi - \alpha^{ee}_{zy} \cos 2\phi &
\alpha^{ee}_{zz}
\end{bmatrix}.
\end{split}
\label{eq:alphamirror}
\end{equation}
Applying \eqref{eq:symcond2} to the first element of \eqref{eq:alphamirror} requires that
\begin{equation}
\alpha^{ee}_{xx} = \alpha^{ee}_{xx} \cos^2 2\phi + \alpha^{ee}_{yy} \sin^2 2\phi + \frac{1}{2}(\alpha^{ee}_{xy} + \alpha^{ee}_{yx}) \sin 4\phi
\end{equation}
for all possible angles $\phi$, which can only be achieved if
\begin{align}
\alpha^{ee}_{xx} &= \alpha^{ee}_{yy}, & \alpha^{ee}_{xy} &= -\alpha^{ee}_{yx}. 
\label{eq:matcond1}
\end{align}
Applying \eqref{eq:symcond2} to the second element of \eqref{eq:alphamirror}, and inserting the restrictions \eqref{eq:matcond1} already obtained gives
\begin{equation}
\alpha^{ee}_{xy} = -\alpha^{ee}_{xy} \cos^2 2\phi - \alpha^{ee}_{xy} \sin^2 2\phi,
\end{equation}
which immediately yields
\begin{equation}
\alpha^{ee}_{xy} = \alpha^{ee}_{yx} = 0.
\end{equation}
Similarly treating the second line of \eqref{eq:alphamirror} imposes the further restrictions
\begin{align}
\alpha^{ee}_{xz} = \alpha^{ee}_{yz} = \alpha^{ee}_{zx} = \alpha^{ee}_{zy} = 0,
\end{align}
while $\alpha^{ee}_{zz}$ is a free parameter.

Repeating the procedure for the restrictions on $\bar{\bar{\alpha}}^{em}$ requires a slight modification to \eqref{eq:symcond1},
\begin{equation}
(\bar{\bar{\alpha}}^{em})' = -T\bar{\bar{\alpha}}^{em}T^{-1},
\label{eq:symcondpsu}
\end{equation}
producing a result almost identical to \eqref{eq:alphamirror}. Applying \eqref{eq:symcond2} and \eqref{eq:symcondpsu}, the key tensor element now reads
\begin{equation}
\alpha^{em}_{xy} = \alpha^{em}_{xy} \cos^2 2\phi - \alpha^{em}_{yx} \sin^2 2\phi - \frac{1}{2}(\alpha^{em}_{xx} - \alpha^{em}_{yy}) \sin 4\phi,
\end{equation}
so 
\begin{align}
\alpha^{em}_{xx} &= \alpha^{em}_{yy}, & \alpha^{em}_{xy} &= -\alpha^{em}_{yx}. 
\end{align}
But now
\begin{equation}
\alpha^{em}_{xx} = -\alpha^{em}_{xx} \cos^2 2\phi - \alpha^{em}_{yy} \sin^2 2\phi - \frac{1}{2}(\alpha^{em}_{xy} + \alpha^{em}_{yx}) \sin 4\phi,
\end{equation}
thus requiring
\begin{equation}
\alpha^{em}_{xx} = \alpha^{em}_{yy} = 0.
\end{equation}
Furthermore,
\begin{align}
\alpha^{em}_{xz} = \alpha^{em}_{yz} = \alpha^{em}_{zx} = \alpha^{em}_{zy} = \alpha^{em}_{zz}= 0.
\end{align}

By similarly treating the other two quadrants of the polarizability tensor, the final symmetry allowed form under the $C_{\infty v}$ point group is obtained,\autocite{barron2004molecular}
\begin{equation}
\begin{bmatrix}
\alpha^{ee}_\perp & 0 & 0 & 0 & -\alpha^{em} & 0\\
0 & \alpha^{ee}_\perp & 0 & \alpha^{em} & 0 & 0\\
0 & 0 & \alpha^{ee}_z & 0 & 0 & 0\\
0 & \alpha^{me} & 0 & \alpha^{mm}_\perp & 0 & 0\\
-\alpha^{me} & 0 & 0 & 0 & \alpha^{mm}_\perp & 0\\
0 & 0 & 0 & 0 & 0 & \alpha^{mm}_z
\end{bmatrix},
\end{equation}
matching \eqref{eq:indtensor} after simplifications to notation enabled by symmetry.

The additional restriction due to $D_{\infty h}$ symmetry is invariance under inversion. Matrix $T$ is particularly simple, with electric quantities transforming as their negative and magnetic as the identity. But as discussed, this fails to hold for the $e^{i\beta z}$ line dipoles, for which \eqref{eq:T} is undefined. Continuing the analysis for inversion symmetry, we apply the equivalent of \eqref{eq:symcond1} to the full tensor,
\begin{equation}
\bar{\bar{\alpha}}' = 
\begin{bmatrix}
\bar{\bar{\alpha}}^{ee} & -\bar{\bar{\alpha}}^{em}\\
-\bar{\bar{\alpha}}^{me} & \bar{\bar{\alpha}}^{mm}
\end{bmatrix},
\end{equation}
so applying the equivalent of \eqref{eq:symcond2} demands that all magnetoelectric quadrants be zero. The final form of the tensor under $D_{\infty h}$ symmetry is 
\begin{equation}
\begin{bmatrix}
\alpha^{ee}_\perp & 0 & 0 & 0 & 0 & 0\\
0 & \alpha^{ee}_\perp & 0 & 0 & 0 & 0\\
0 & 0 & \alpha^{ee}_z & 0 & 0 & 0\\
0 & 0 & 0 & \alpha^{mm}_\perp & 0 & 0\\
0 & 0 & 0 & 0 & \alpha^{mm}_\perp & 0\\
0 & 0 & 0 & 0 & 0 & \alpha^{mm}_z
\end{bmatrix},
\end{equation}
matching \eqref{eq:alphaTMTE}.

\section{Energy Conservation}
\label{sec:sipekran}
Energy conservation imposes restrictions on the polarizability tensor in the form of the Sipe-Kranendonk relations, also known as the optical theorem.\autocite{sipe1974macroscopic,belov2003condition,sersic2011magnetoelectric,strickland2015dynamic} These are a direct consequence of the balance between energy extinguished from the incident field and the energy radiated by lossless point dipoles. These relations were derived by Strickland et al.\ for the magnetoelectric formulation $\bar{\bar{\alpha}}$, but can also be derived for the TM/TE formulation \eqref{eq:splitalpha}, with restrictions individually applicable to each tensor, $\bar{\bar{\alpha}}^\textrm{TM}$ and $\bar{\bar{\alpha}}^\textrm{TE}$, of \eqref{eq:alphaTMTE}.

The total radiated energy per unit length is given by an integral enclosing the cylinder of the Poynting vector,
\begin{equation}
P_\textrm{rad}/l = \frac{1}{2Z_0} \real\int \bv{E}\times\bv{H}^* \cdot d\bv{n}.
\end{equation}
This can be evaluated directly in terms of the line dipoles \eqref{eq:linedipoles},\autocite{strickland2015dynamic}
\begin{equation}
\begin{split}
P_\textrm{rad}/l =\ &\frac{k}{16Z_0}[2k_\perp^2 (|p_z|^2+|m_z|^2) + (k^2+\beta^2)(|p_x|^2+|p_y|^2+|m_x|^2+|m_y|^2)\\
&+ 2k\beta(p_xm_y^* + p_x^*m_y - p_ym_x^* - p_y^*m_x)].
\label{eq:prad}
\end{split}
\end{equation}
Meanwhile, the energy extinguished from the incident field is given by
\begin{equation}
P_\textrm{ext}/l = \frac{k}{2Z_0}\imag[\bv{p}\cdot\bv{E}^* + \bv{m}\cdot\bv{H}^*].
\label{eq:pext}
\end{equation}

To express results directly in terms of elements of the polarizability tensor, we substitute either $\bar{\bar{\alpha}}^\textrm{TM}$ or $\bar{\bar{\alpha}}^\textrm{TE}$ of \eqref{eq:alphaTMTE} into \eqref{eq:prad} and \eqref{eq:pext}, giving
\begin{gather}
\begin{split}
P_\textrm{rad}/l =\ &\frac{k}{16Z_0}[2k_\perp^2 (|\alpha^{ee}_z|^2|E_z|^2+|\alpha^{mm}_z|^2|H_z|^2) + (k^2+\beta^2)\{|\alpha^{ee}_\perp|^2(|E_x|^2+|E_y|^2)+|\alpha^{mm}_\perp|^2(|H_x|^2+|H_y|^2)\}\\
&+ 4k\beta\real(\alpha^{ee}_\perp\alpha^{mm*}_\perp E_xH_y^* - \alpha^{ee}_\perp\alpha^{mm*}_\perp E_yH_x^*)],
\end{split}\\
P_\textrm{ext}/l = \frac{k}{2Z_0}\imag[\alpha^{ee}_z|E_z|^2 + \alpha^{mm}_z|H_z|^2 + \alpha^{ee}_\perp(|E_x|^2+|E_y|^2) + \alpha^{mm}_\perp(|H_x|^2+|H_y|^2)].
\end{gather}
Here the symbol $\alpha^{ee}_\perp$, for example, can refer to either $\alpha^{ee,\textrm{TM}}_\perp$ or $\alpha^{ee,\textrm{TE}}_\perp$, depending on which of \eqref{eq:alphaTMTE} was substituted.

The Sipe-Kranendonk relations are then derived by equating term-by-term. Treating first the TM case, contributions from the $z$-components can be separated and the fields canceled to derive
\begin{equation}
4\imag(\alpha^{ee,\textrm{TM}}_z) = k_\perp^2|\alpha^{ee,\textrm{TM}}_z|^2.
\label{eq:skeez}
\end{equation}
The in-plane contribution to $P_\textrm{rad}$ has cross terms that depend on both $\bv{E}$ and $\bv{H}$, while $P_\textrm{ext}$ does not. This can be resolved by substituting the characteristic ratios between field component pairs $E_x/H_y$ and $E_y/H_x$, specific to each polarization. These four field components yield two equations, which are identical as a result of symmetry,
\begin{equation}
8\imag(\beta^2\alpha^{ee,\textrm{TM}}_\perp + k^2\alpha^{mm,\textrm{TM}}_\perp) =
k_\perp^2k^2|\alpha^{mm,\textrm{TM}}_\perp|^2 - k_\perp^2\beta^2|\alpha^{ee,\textrm{TM}}_\perp|^2 + 2k^2\beta^2|\alpha^{ee,\textrm{TM}}_\perp + \alpha^{mm,\textrm{TM}}_\perp|^2.
\label{eq:sktm}
\end{equation}
The procedure can be repeated for the opposite polarization completing the Sipe-Kranendonk relations
\begin{gather}
4\imag(\alpha^{mm,\textrm{TE}}_z) = k_\perp^2|\alpha^{mm,\textrm{TE}}_z|^2,\\
8\imag(k^2\alpha^{ee,\textrm{TE}}_\perp + \beta^2\alpha^{mm,\textrm{TE}}_\perp) =
k_\perp^2k^2|\alpha^{ee,\textrm{TE}}_\perp|^2 - k_\perp^2\beta^2|\alpha^{mm,\textrm{TE}}_\perp|^2 + 2k^2\beta^2|\alpha^{ee,\textrm{TE}}_\perp+\alpha^{mm,\textrm{TE}}_\perp|^2.
\label{eq:skte}
\end{gather}

\printbibliography
\end{document}